\documentclass[aps,twocolumn,nofootinbib,showpacs,floatfix]{revtex4}

\usepackage[centertags]{amsmath}
\usepackage{amsfonts}
\usepackage{graphicx}
\usepackage[hypertex]{hyperref}
\usepackage{psfrag}

\newcommand{\bea}{\begin{eqnarray}}
\newcommand{\eea}{\end{eqnarray}}
\newcommand{\be}{\begin{equation}}
\newcommand{\ee}{\end{equation}}

\newcommand{\Lo}{\Lambda_{0}}

\def\spose#1{\hbox to 0pt{#1\hss}}
\def\ltapprox{\mathrel{\spose{\lower 3pt\hbox{$\mathchar"218$}}
\raise 2.0pt\hbox{$\mathchar"13C$}}}
\def\gtapprox{\mathrel{\spose{\lower 3pt\hbox{$\mathchar"218$}}
\raise 2.0pt\hbox{$\mathchar"13E$}}}
\def\inapprox{\mathrel{\spose{\lower 3pt\hbox{$\mathchar"218$}}
\raise 2.0pt\hbox{$\mathchar"232$}}}
 
\newcommand{\ve}[1]{\ensuremath{\mathbf{#1}}}
\newcommand{\D}[1][ ]{\ensuremath{\mathrm{d}^{#1} }}
\newcommand{\tdyn}{\ensuremath{\tau_\text{dyn}}}

\begin{document}

\title{Gravitational dynamics of an infinite shuffled lattice:
early time evolution and universality of non-linear correlations} 

\author{T. Baertschiger}
\affiliation{Dipartimento di Fisica, Universit\`a ``La Sapienza'',
P.le A. Moro 2, I-00185 Rome, Italy,\\ \& ISC-CNR, Via dei Taurini 19,
I-00185 Rome, Italy.}
\author{M. Joyce}
\affiliation{Laboratoire de Physique Nucl\'eaire et de Hautes Energies, 
UMR 7585, \\
Universit\'e Pierre et Marie Curie --- Paris 6, 75252 Paris Cedex 05, France.}
\author{F. Sylos Labini}
\affiliation{  "E. Fermi'' Center, Via Panisperna 89 A, Compendio del
Viminale, I-00184 Rome, Italy,\\ \& ISC-CNR, Via dei Taurini 19,
I-00185 Rome, Italy.}
\author{B. Marcos}
\affiliation{  "E. Fermi'' Center, Via Panisperna 89 A, Compendio del
Viminale, I-00184 Rome, Italy,\\ \& ISC-CNR, Via dei Taurini 19,
I-00185 Rome, Italy.}

\begin{abstract}    
In two recent articles a detailed study has been presented of the 
out of equilibrium dynamics of an infinite system of self-gravitating
points initially located on a randomly perturbed lattice. In this
article we extend the treatment of the early time phase during which strong
non-linear correlations first develop, prior to the onset of
``self-similar'' scaling in the two point correlation function. We
establish more directly, using appropriate modifications of the
numerical integration, that the development of these correlations can
be well described by an approximation of the evolution in two phases:
a first perturbative phase in which particles' displacements are small
compared to the lattice spacing, and a subsequent phase in which
particles interact only with their nearest neighbor. For the range of
initial amplitudes considered we show that the first phase can be well
approximated as a transformation of the perturbed lattice
configuration into a Poisson distribution at the relevant scales. This
appears to explain the ``universality'' of the spatial dependence of
the asymptotic non-linear clustering observed from both shuffled
lattice and Poisson initial conditions.
\end{abstract}    
\maketitle    
\date{today}    

\twocolumngrid  

\section{Introduction}

Structure formation in the universe is currently addressed primarily 
using numerical simulations of purely self-gravitating particle
systems, with initial configurations generated by displacing the
particles slightly from a perfect lattice (see e.g.
\cite{efstathiou_88, couchman}). The physics of the strongly
non-linear regime of the observed evolution is, in detail, very poorly
understood.  Progress in understanding would be very useful in
providing analytical guidance for numerical simulations, and in
particular better control on their precision in representing the
relevant continuum limit. In a series of recent papers 
\cite{sl1,sl2} we have studied a reduced version of the full 
cosmological problem, considering a very simple class of randomly 
perturbed lattices as
initial conditions, and evolution in a static universe\footnote{See also \cite{Baertschiger:2004tx, maurizio} for 
earlier studies of evolution from these initial conditions.}.

One of our primary results is that, despite the 
simplifications, the system we study has qualitative behavior very
similar to that observed in the more complex cosmological simulations
(correlated perturbations in initial conditions, expanding universe).
Notably the evolution is clearly ``hierarchical'' (i.e. structures
build up at successively larger scales driven by the linearized fluid
theory growth of the initial perturbations), and asymptotically
``self-similar'' (i.e. the time dependence of the two point correlation
function is given by a simple scaling of the spatial variable which 
can be inferred from the linearized fluid 
theory)\footnote{To avoid any possible confusion we note that both 
these terms are used here with meanings different to those commonly 
ascribed to them in  condensed matter physics. In the latter context 
both are associated with invariance properties of the spatial 
correlations under spatial rescalings (see e.g. \cite{book}).
Such properties are not implied by their use in the present context.}. 
The {\it functional form} of the spatial dependence 
of the non-linear correlation function is, on the other hand, just as in the
cosmological simulations, a fundamental quantity characterizing the
bresults which is determined numerically, but not currently understood
(i.e. not predicted analytically or semi-analytically). We have noted
in \cite{sl1,sl2}, however, that this form emerges, to a good
approximation, in our simulations {\it prior} to the asymptotic
bscaling regime, in the preceding transient phase in which strong
non-linear correlations first develop. In this paper we extend and
detail our analysis of this phase. We show in greater detail that the
emergence of the observed non-linear two body correlations can be very
well approximated by modeling the evolution as constituted of two
subsequent phases, with an abrupt matching from one to the other.
During the first phase the particles evolve as described by a
perturbative analytical approximation we have introduced and studied
in \cite{joyce_05, marcos_06}; in the second phase the particles
evolve under a force coming solely from their nearest neighbor (NN).

We relate our work here to previous work along similar lines
concerning evolution from Poissonian initial 
conditions \cite{bottaccio3, maurizio, Baertschiger:2004tx}. In this 
case it has been shown explicitly \cite{Baertschiger:2004tx} that 
the emergence of the 
first strong non-linear correlations can be very well accounted for by
an approximation, at the relevant early times, in which the full
gravitational force on each particle is truncated to that due only to
its initial NN\footnote{The importance of NN interactions at early times
starting from Poisson initial conditions has also been discussed previously
by Saslaw. See \cite{saslaw2000} and references therein.}. In the case 
of ``shuffled lattice'' (SL) initial
conditions, which we consider in this work, such an approximation is
not generically good: when the typical displacement of a particle is
small compared to the lattice spacing, the high degree of symmetry
gives that the force on a typical particle is the sum of comparable
contributions from many particles. In this regime, however, we can
describe the evolution very well by a simple perturbative
approximation, which has been developed fully in \cite{joyce_05,
marcos_06}.  This latter approximation breaks down, roughly, when
particles start to approach one another, which is precisely when one
expects a NN approximation for the force may become
appropriate.  We show here that this is indeed the case, and that an
abrupt switch between the two phases gives a very good approximation
to the evolution. Further this model allows us to explain the fact
that the observed form of the non-linear correlations in our
simulations is independent of the amplitude of the initial shuffling,
and the same as that observed from Poisson initial conditions. This is
the case because, for the range of amplitude of the initial
perturbations we use, the evolution in the first phase brings the
system to a distribution with correlation properties which, at the
relevant scales, are essentially those of a Poisson distribution.

The main interest of our results here is that they give 
a semi-analytical understanding of the origin of the form
of the observed non-linear two point correlations for this 
class of initial conditions, which are qualitatively similar 
to those used in cosmological type simulations. As remarked 
in \cite{sl1,sl2} the form of this early time correlation
function coincides, to a very good approximation, with that 
which is also observed in the asymptotic scaling regime
attained by the system at longer times. This suggests strongly 
(but does not prove) that the physical mechanism leading to
the former correlations, which we identify here, is also
that which gives rise to the latter correlations. In the context
of cosmological simulations such a conclusion, if appropriate 
also for that case, would be very important for the following reason.
In this context the results derived from the numerical simulations 
which are physically relevant  are those which are representative 
of the Vlasov-Poisson (VP) limit of the simulated particle system. 
The mechanism we describe here for the generation of the non-linear 
correlations is, on the other hand, clearly {\it not} representative 
of this limit: 
the effect of interactions with single NN particles are precisely
of the kind which are discarded in the VP limit, which is a 
mean-field approximation. Therefore, if the form of the non-linear
correlations in the long time evolution turns out actually to be
determined in such a phase, this form would not be representative,
as required, of the VP limit. We discuss this point a little further 
in our conclusions, and suggest numerical tests which could be performed 
to determine whether the long-time behavior is indeed linked to
the early time mechanism we study here. 

The paper is organized as follows. In Sec.~\ref{sec2} we discuss some
of the relevant properties of the initial conditions of our
simulations.  In the next section we give the details of the
simulations considered here and summarize briefly the main relevant
results of \cite{sl1,sl2}.  In Sec.~\ref{sec3} we present in detail the
two-phase model which captures the essential elements of the formation
of the first non-linear correlations. We give results here also of
numerical simulations.  Finally in Sec.~\ref{sec4} we discuss the
results and draw our main conclusions.

\section{Statistical properties of initial conditions}
\label{sec2}

As in \cite{sl1} we study evolution from initial conditions
in which the particles are at rest and located at the sites of a perfect simple
cubic lattice subjected to random uncorrelated displacements.
We adopt the same notation, denoting 
by $p(\ve u)$ the probability density
function (PDF) for the displacements, and by $\ve u (\ve R)$ the displacement
of the particle originally at lattice site $\ve R$. The variance
of the PDF is denoted by $\Delta^2$, and the dimensionless 
variance $\delta^2 \equiv \Delta^2/\ell^2$, where $\ell$ is 
the lattice spacing. The parameter $\delta$ we refer to as  
{\it the normalized shuffling parameter}. As discussed in
\cite{sl1,sl2}, for the case of purely gravitational  
interactions, the system is completely characterized, in the infinite
volume limit, by the single parameter $\delta$.
The Poisson
distribution corresponds to the limit in which each
particle's position is completely randomized in the infinite volume,
i.e., $\delta=\infty$.

The precise details of the different initial conditions of which the
evolution is studied numerically below, are summarized in
Tab.~\ref{tab:sl_summary}. The PDF $p(\ve u)$ used for generating the
displacements is constant in a cube of side $2 \Delta$ around the
origin (and with sides parallel to the axes of the lattice), and zero
elsewhere. 
The (arbitrary) choice of units is as in \cite{sl1}, giving
that 
the {\it dynamical time} $\tau_{dyn} \equiv 1/\sqrt{4\pi G \rho_0}$ 
is equal (where $\rho_0$ is the mass density).

The first four simulations (SL64 to SL16) are the same ones 
analyzed in \cite{sl1}. As explained there (see Sec. III of \cite{sl1}) the
values of $\delta$ have been chosen so that, in our units of length,
$\delta^2 \ell^5$ is constant. This gives (see \cite{sl1} for details)
an amplitude of the power spectrum at small $k$ (i.e. $k \ll
\ell^{-1}$ for $\delta < 1$) 
which is equal in all simulations. With time in units of 
$\tau_{dyn}$ this means that, in the long wavelength fluid limit, the systems
are identical initially, and evolve identically.  SL64b differs only
from SL64 in the value of $\delta$, i.e., they are two simulations
with identical values of the parameters characterizing the finite
numerical representation of the infinite systems, but with different
$\delta$.

\begin{table}
\begin{ruledtabular}
\begin{tabular}{cccccccc}
Name & $N^{1/3}$ & $L$ & $\ell$ & $\Delta$ & $\delta$ & $m/m_{64}$ \\ 
\hline 
SL64 & 64 & 1& 0.015625& 0.015625 & 1       & 1     \\ 
SL32 & 32 & 1&0.03125  & 0.0553   & 0.177   & 8     \\ 
SL24 & 24 & 1&0.041667 & 0.00359  & 0.0861  & 18.96 \\ 
SL16 & 16 & 1&0.0625   & 0.00195  & 0.03125 & 64    \\ 
\hline 
SL64b  & 64 & 1& 0.015625& 0.0012  & 0.0768    & 1    \\ 
\hline 
P64  & 64 & 1& 0.015625& $\infty$  & $\infty$      & 1    \\ 
\end{tabular}
\end{ruledtabular}
\caption{Details of the initial conditions studied in this paper,
and the numerical parameters used in the simulations. 
$N$ is the number of particles in the cubic box of side $L$,
and $m$ is the particle mass.
\label{tab:sl_summary} }
\end{table}

To characterize the correlation properties of the distributions
we will use the same quantities as in \cite{sl1,sl2}: the
reduced two point correlation function $\xi(\ve r)$,
the power spectrum $P(\ve k)$ (which is related to $\xi(\ve r)$
by a Fourier transform\footnote{The power spectrum is the
Fourier transform of $\tilde \xi(\ve r)$, which differs from 
what we refer to here as the ``correlation function'' by a
delta function singularity at $r=0$. See \cite{sl1,sl2} or
\cite{book}.}), the NN PDF $\omega(r)$.
We refer the reader to \cite{sl1} for the precise definitions
of these quantities.
We will also consider the stochastic properties of the force,
which we characterize using $P(F)$, the PDF for the modulus of
the force $F$. 

While $\delta=\infty$ corresponds exactly to the Poisson distribution,
one expects any SL with $\delta \geq 1$ to approximate the
correlation properties of a Poisson distribution up to 
a scale of order $\Delta=\delta \ell$. Indeed, if $\delta \geq 1$, the
effect of the short distance exclusion of the underlying lattice should
disappear and the particles are, to a good approximation, randomly placed
in a volume of order $\Delta^3$. This can be seen explicitly for the power 
spectrum, of which the exact analytical expression may be written 
\cite{book,sl1} in the form 
\begin{equation}
\label{eq_pk} 
P(\ve k)= \frac{1}{n_0} + |\tilde{p} (\ve k) |^2 A(\ve k)
\end{equation}
where $n_0$ is the mean particle density and $\tilde{p} (\ve k)$ is
the characteristic function of $p (\ve u)$ (i.e its Fourier transform
normalized so that $\tilde{p} (0)=1$). The function $A(\ve k)$ depends
only on the initial unperturbed lattice distribution (and not on the
shuffling). For any simple form of the PDF for the shuffling (such as
the top-hat one considered here and in \cite{sl1}, or, e.g., a
Gaussian PDF as used in \cite{sl2}) $\tilde{p} (\ve k)$ decreases
toward zero for $k \Delta > 1$, giving that the power spectrum tends
to the Poissonian value (given by $1/n_0$). Thus for wave-numbers $k$
larger than of order $1/(\delta \ell)$ the power spectrum converges 
to this Poissonian
behavior. We refer the reader notably to Fig. 2 of \cite{sl1}, which
show the power spectrum for the four initial conditions SL16, SL24,
SL32 and SL64.

\subsection{Nearest neighbor distribution}
\label{Nearest neighbor distribution}

We will consider below often the NN PDF, and it is useful to
know its form in the initial conditions just described. For the
Poissonian limit $\delta=\infty$ it is straightforward to show 
analytically \cite{book} that it is 
\begin{equation}
\omega_P (r)= 4\pi n_0 r^2 \exp \left(-\frac{4}{3}\pi n_0 r^3 \right)
\label{poisson-NNpdf}
\end{equation}
which gives an average distance between NN  
$\Lo\approx 0.55 n_0^{-1/3}=0.55 \ell$.
Since the NN distribution characterizes the small scale properties 
we expect, following our discussion above, that this expression will 
be a good approximation for $\delta \geq 1 $. For $\delta \ll 1$, on 
the other hand, one may show\footnote{The derivation of the expression
given is straighforward, but tedious. For a given shuffling of a
particle and its six NN, one must determine exhaustively 
the different combinations, and associated probabilities, which 
lead to a given NN separation. The approximation $\delta \ll 1$
is used in taking the inter-particle separations to linear order
in $\delta$.} that
\be
\ell \omega(r) \approx \frac{1}{(2\delta)^9} f \left(\frac{r}{\ell}-1\right)
\label{eq:omegasmall}
\ee
where
\be
f(x)= \begin{cases}
\delta(2\delta + x)\left[2\delta x^2 + 8\delta^2 x \right]^2  
& \text{if } x \in [-2\delta,-\delta] \\
\left[ \delta^2 - x^2 + \delta x \right] \\
\times \left[ 
2\delta x^2 +\frac{4}{3}
  \delta^3-4\delta^2x+\frac{4}{3}x^3\right]^2 & \text{if }
  x \in \ ]-\delta,0] \\
\frac{16}{9}(\delta - x)^8 & \text{if } x \in \ ]0,\delta]
    \\
0 & \text{otherwise},
\end{cases}
\nonumber
\ee
corresponding to average distance between NN 
$\Lo = \ell - (86827/80640) \Delta$.  

\begin{figure}
  \includegraphics[width=0.5\textwidth]{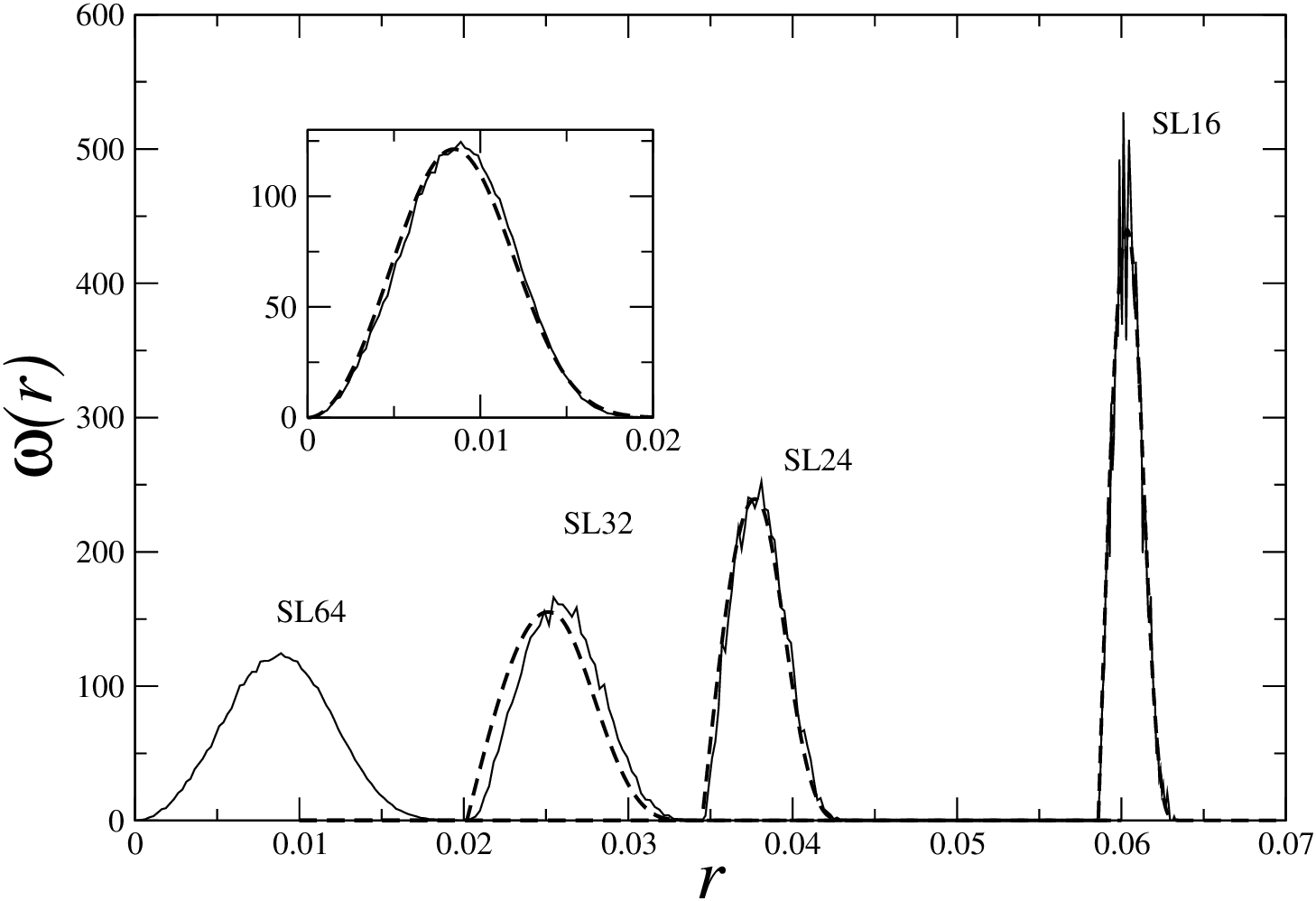}
\caption{NN PDF of the SL considered in this
paper (the name of each SL is indicated above the corresponding
curve). For SL32, SL24 and SL16, the function~\eqref{eq:omegasmall}
is shown for comparison. 
In the insert panel we show enlarged the SL64 (SL) case together
with the behavior of $\omega_p(r)$ for a Poisson distribution (P) with
same number density [i.e. Eq.\eqref{poisson-NNpdf}]. 
\label{omegat0}}
\end{figure}

In Fig.~\ref{omegat0} we show the behavior of $\omega(r)$ for most of 
the SL studied here. For the SL with very small shuffling 
---~SL16 or SL24~--- this PDF is strongly peaked around the average 
distance between NN (which is approximately equal to $\ell$), and 
in very close agreement with the analytical approximation given
by Eq.~(\ref{eq:omegasmall}). For SL32 (with $\delta=0.177$) a small
discrepancy with this approximation is visible, while for SL64 
(with $\delta=1$) the NN PDF is, as expected, in very good agreement 
with that in the Poisson case given by Eq.~(\ref{poisson-NNpdf}). 

\subsection{Force distribution}
\label{Force distribution}

\begin{figure}
\includegraphics[width=0.5\textwidth]{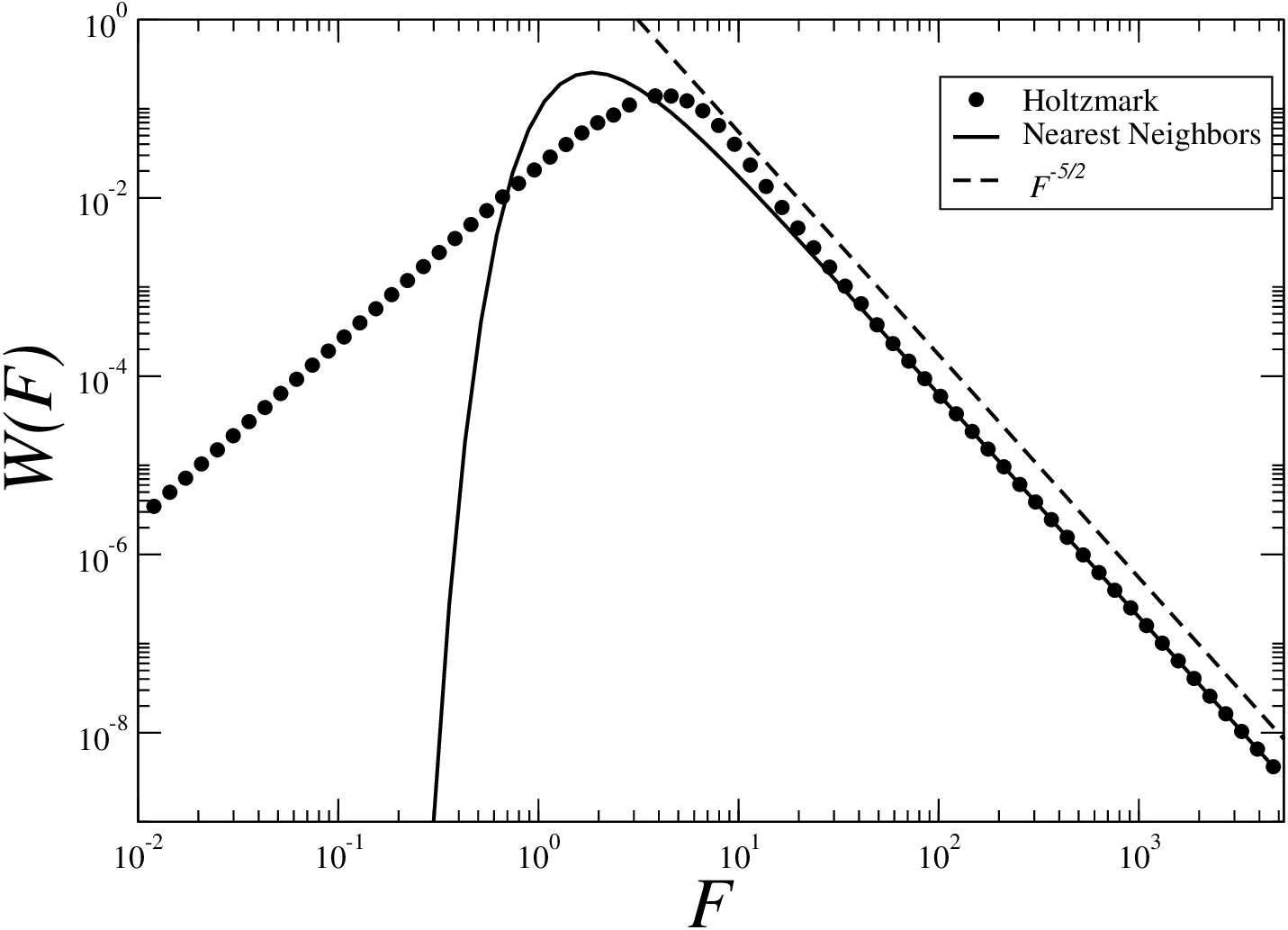}
\caption{Holtzmark
distribution and the PDF inferred if only NNs contribute,
i.e., $W_{NN}(F)dF=\omega (r) dr$. The agreement is very good in the strong
field limit where $W(F) \sim F^{-5/2}$. At weak field the PDF due to
the NN has a sharp cut-off while the Holtzmark distribution shows
a more gentle decrease (see discussion in \cite{book,gabrielli_06}).
\label{forceHNN}}
\end{figure}

The PDF of the modulus of the force $W(F)$ is a useful quantity in our
analysis. Notably if the forces on particles are dominated by that 
coming from their NN the simple relation 
\begin{equation}
W(F)dF=\omega (r)dr
\label{NNdomination}
\end{equation}
must hold. For the case of a
Poisson distribution the analytical expression for $W(F)$ was first
given by Chandrasekhar \cite{chandra43}. It is proportional to the 
so-called Holtzmark
distribution (see \cite{book} for the explicit result and a simple
derivation). In Fig.~\ref{forceHNN}
we show a plot of the full
(Holtzmark) distribution, and the PDF inferred if only NNs
contribute $W_{NN}(F)dF=\omega (r) dr$. 
The domination of the NN in the force is clearly seen at stronger
values of the force. The relation is not valid at weaker values of the
force as these correspond to the (rare) particles for which the force
picks up comparable (and possibly canceling) contributions from more
than one particle. Note that the tail of the PDF at large $F$ decays in
proportion to $F^{-5/2}$, which means that the variance (i.e. second
moment) of the PDF is infinite.

In \cite{gabrielli_06} we have studied in detail the statistical properties
of the force in an SL. As one might expect, one can show that the force
PDF is very well approximated by that of a Poisson distribution when 
the typical displacement is larger than the inter-particle spacing, i.e.,
$\delta > 1$. At small values of the displacements, on the other hand,
the force PDF is very different to that in a Poisson distribution,
decaying much more rapidly at large values of the force: the strong
forces due to NN are completely absent as the typical particle feels
a comparable effect from its {\it six} NN when the configuration is
close to a perfect simple cubic lattice. More precisely, for 
a top-hat PDF of the displacements (as used here), the functional
behavior of the PDF at large $F$ changes qualitatively, from exponential
decay to a $F^{-5/2}$ power law decay, at $\delta=0.5$. For 
$\delta \gtapprox 0.5$ the amplitude of the latter tail
is lower than that in the Poisson, with this difference becoming
negligible as $\delta$ increases to of order unity. 

\section{Numerical simulations} 
\label{sec3} 

In this section we first report results of numerical simulations
in which the initial conditions 
given in Tab.~\ref{tab:sl_summary} are evolved under the
mutual self-gravity of all particles. As such simulations have already
been reported in detail in \cite{sl1,sl2} (see also 
\cite{Baertschiger:2004tx, bottaccio3} for Poisson
initial conditions and some SL), we restrict 
ourselves to a very brief summary with an
emphasis on the points which are relevant here. We then report results
of a new set of simulations designed to validate our model of the early
time evolution by a direct numerical integration of the appropriate two
phase approximation. 

\subsection{Full gravity} 

As in \cite{sl1,sl2} we have used the publicly available code  
\textsc{Gadget} \cite{gadget,gadget_paper} to evolve the system under 
gravity, modified only by a small scale regularization of the
potential below $r=\varepsilon$. This softening parameter
$\varepsilon$ is taken here in all simulations to be $\varepsilon =
0.00175L$ (i.e. in all simulations significantly smaller
than the initial average distance between NN). We have 
performed the same checks as discussed in \cite{sl1,sl2} 
for the independence of our results to this choice.
%

\begin{figure}
\scalebox{0.35}{\includegraphics*{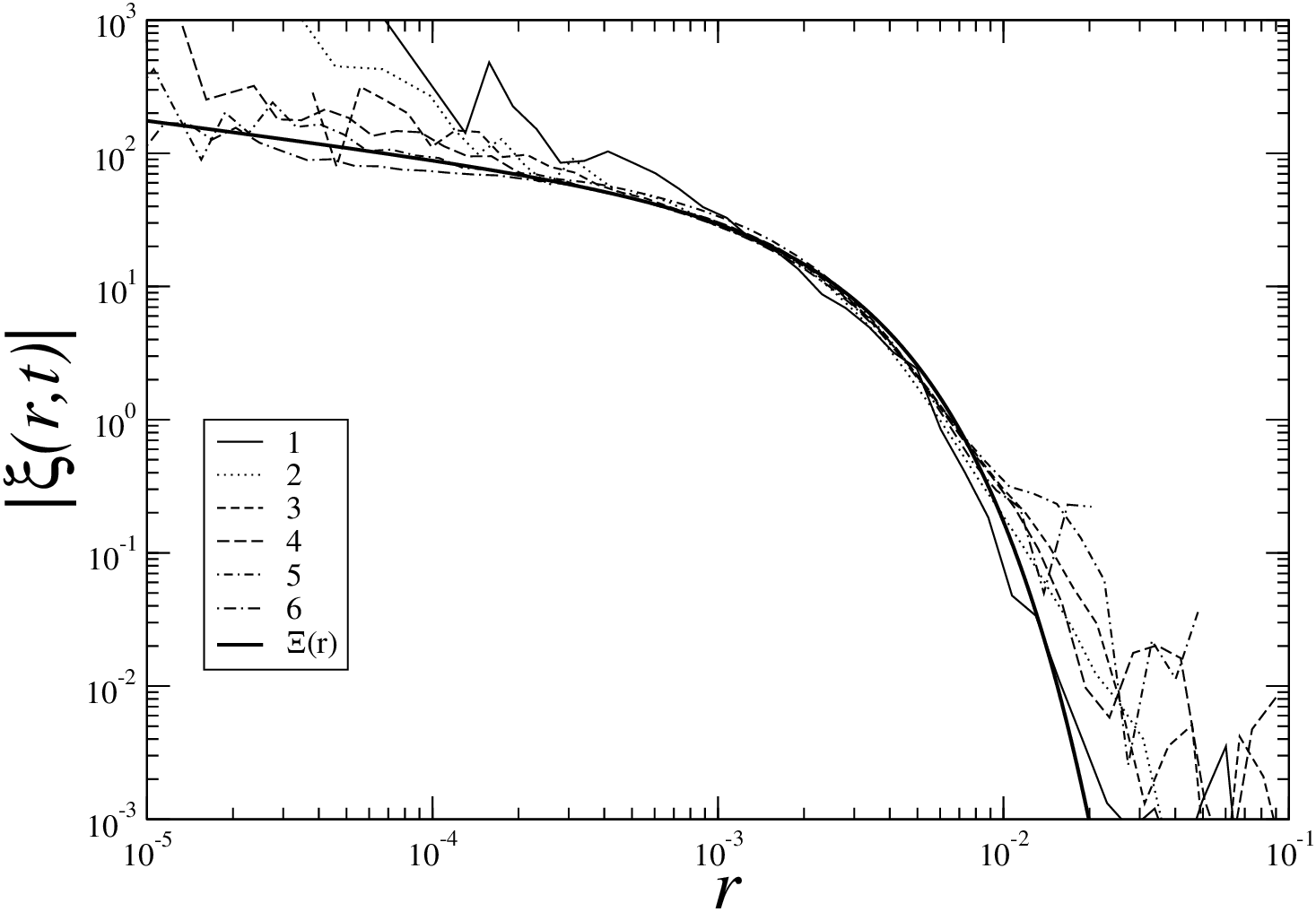}}
\caption{``Collapse plot'' of $\xi(r,t)$ (absolute value)
for the P64 initial conditions: for each time $t>1$ we 
have rescaled $r$ so that $\xi(r,t)=1$ at $r_0$,
where $\xi(r_0,t=1)$. The behavior of the numerical
fit given in \cite{sl1} is also shown for comparison.
\label{xi.p64.R}}
\end{figure}

\begin{figure}
\scalebox{0.35}{\includegraphics*{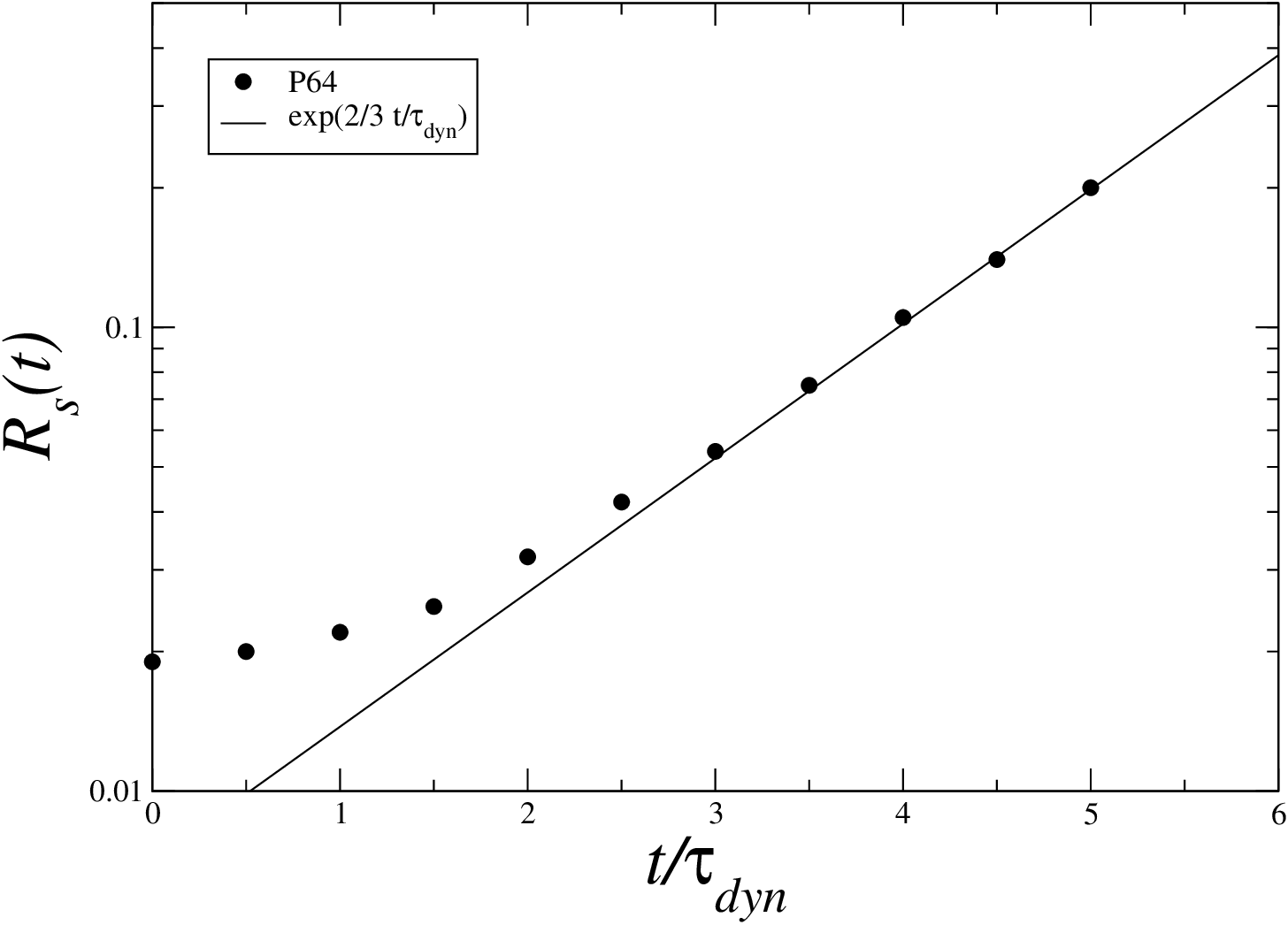}}
\caption{Evolution of the function $R_s(t)$ in P64 
(points) compared with its prediction from linear fluid 
theory, $R_s(t)\propto \exp[ (2/3)
t/\tdyn]$.  }
\label{xi.p64.RS(t)}
\end{figure}

In \cite{sl1} we have found that the SL initial conditions
considered here lead, from the time significant non-linear correlation 
first develops at small scales, to an evolution in which the 
correlation function can be approximated by
\begin{equation}
\xi(r,t) \approx \Xi\left( r/R_s(t) \right)\;, 
\label{eq:ss}
\end{equation}
where $R_s(t)$ is a time dependent length scale, and a simple
functional fit to $\Xi(r)$ is given in \cite{sl1}.
For sufficiently
long times --- after a transient time of which the duration 
increases as  the value of $\delta$ decreases --- $R_s(t)$ 
follows very well the behavior
predicted by a simple analysis based on the linearized equations for
the system approximated as a pressure-less self-gravitating fluid.
As described in \cite{sl1}, for such a system with power 
spectrum of density fluctuations at small wave-numbers $P(k) \sim
k^n$ one obtains:
\be
\label{rs_sl} 
R_s(t) \propto \exp\left(\frac{2}{3+n}\frac{t}{\tdyn}\right) \;.
\ee
As reported in \cite{sl1} SL initial conditions indeed produce the
predicted asymptotic time dependence, corresponding to $n=2$
in this formula.
In Fig.~\ref{xi.p64.R} we show the same analysis 
of the evolved Poisson initial conditions P64,
using the same numerical fit to $\Xi(r)$ as found
in \cite{sl1} for the SL initial conditions.
In Fig.~\ref{xi.p64.RS(t)} 
is shown the associated temporal evolution of $R_s(t)$, and a fit to
the theoretical fluid behavior, given by Eq.~(\ref{rs_sl}) with
$n=0$. The agreement, after a short transient, is as good as 
that observed in \cite{sl1} for the SL.

The conclusion which follows is thus that, while the temporal
behavior of the scaling depends on the value of $\delta$, the
functional form of the spatial dependence in the non-linear
correlation function appears to be the same for all initial
conditions. This is what we refer to as ``universality'' of
the non-linear correlations in this context.  

We note that, just as underlined for the SL initial
conditions in \cite{sl1,sl2}, it is also true for
the Poisson initial conditions that the spatio-temporal scaling of the
two point correlation function, given by Eq.~(\ref{eq:ss}), is a good 
approximation well before the asymptotic scaling behavior, given
by Eq.~(\ref{rs_sl}), sets in. While the dynamical model we present 
below is valid only in this first phase (i.e. that prior to the
asymptotic regime), it is thus natural to hypothesize that the 
form of the asymptotic correlation function is in fact 
determined in this phase. We will discuss this hypothesis
further in our conclusions, and in particular, how it could 
be tested for.

The second essential result about the development of non-linear
correlations which we recall is the following. In all these
simulations (both Poisson \cite{Baertschiger:2004tx} and SL
\cite{Baertschiger:2004tx,sl1}) we observe that the relation  
\begin{equation} 
\label{eq:omega1} 
\omega(r) \, \D r = \left( 1 - \int_0^r\omega(s)\, \D s \right) \cdot
 4\pi r^2  n_0 \left( 1 + \xi (r) \right) \D r  \;,
\end{equation}
holds to a very good approximation, from the time that significant 
non-linear correlations first develop until a time of order a 
dynamical time later.
As explained in \cite{sl1} (see also \cite{book}), it is
valid if all but the two point correlations are trivial. It is 
thus natural
to interpret its observed approximate validity for the correlations
which develop in the first phase of non-linearity to indicate that
these correlations develop predominantly {\it as a result} of the two body
clustering of NN pairs of particles. For the case of the Poisson
initial conditions it has been shown explicitly in
\cite{Baertschiger:2004tx} that this interpretation is correct:
by integrating from the initial conditions with only forces
between initial NN pairs, the evolution of correlation is well
described up to approximately one dynamical time when non-linear
correlations have developed up to a scale of order $\ell$.  

\subsection{Two phase model evolution} 

For SL initial conditions, with small $\delta$, the approximation
of forces as NN dominated is, as we have discussed above, not valid at 
early times. In this limit, however, we have developed in 
\cite{joyce_05,marcos_06} an analytical perturbative approach,
which at linear order gives a very good approximation to the
dynamical evolution. The treatment involves simply a Taylor expansion
of the force between particles in their {\it relative} displacements
from their initial lattice positions $\ve R$, and thus breaks down 
when the latter become equal to the initial separation of the 
particles. In \cite{marcos_06} the precision of the linearized
approximation has been explored in detail for the SL initial
conditions (and others). For the evolution of the average relative 
distance between NN, the approximation turns out to be very good 
until this quantity becomes quite close to the initial lattice 
spacing, i.e., until when many particles come close to their NN. We refer
to this approximation as {\it particle linear theory} (PLT) as 
it is simply a generalization for particles of an analogous
standard treatment for the self-gravitating fluid (in the Lagrangian
formalism, leading to the so-called Zeldovich approximation).
We do not detail further the implementation of this PLT 
approximation here as a succinct summary may be found in \cite{sl2}, and 
a very complete discussion in \cite{marcos_06}.

\begin{figure}[tpd] 
\scalebox{0.35}{\includegraphics*{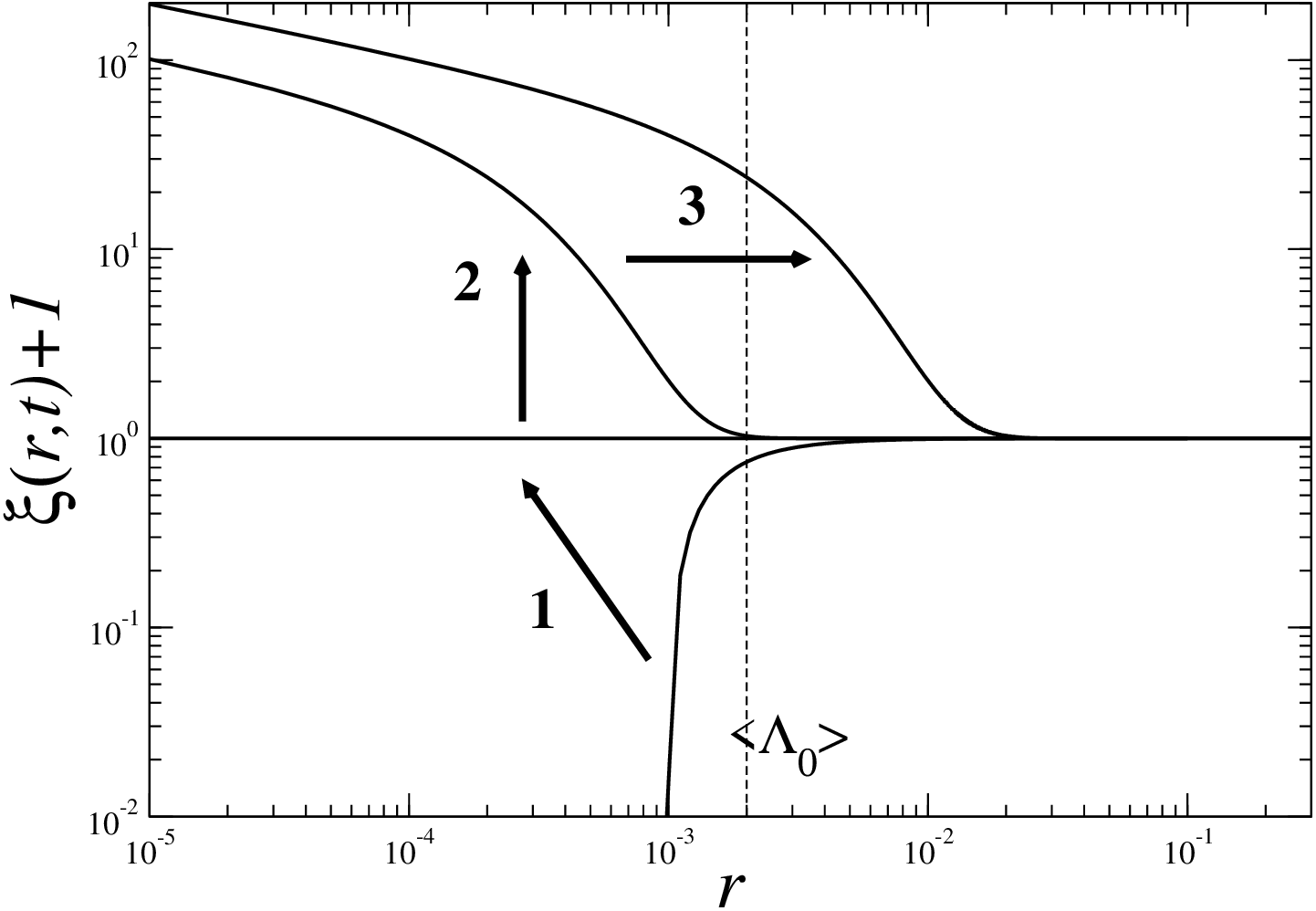}} 
\caption{Schematic representation of the evolution of two point
correlations [specifically $\xi(r)+1$] from SL initial conditions
with small $\delta$. During the first phase (1) the initial 
anti-correlation (i.e. exclusion) at small scales is destroyed,
as particles evolve in the regime described by PLT. In the second 
phase (2) large positive correlations are created at small scales,
up to roughly the initial lattice spacing. The forces responsible 
for these correlations are predominantly those of exerted by NN
pairs on one another. In the subsequent evolution (3), when 
this approximation is no longer valid, the regime of positive
correlations grows in a self-similar way, larger and larger scales 
becoming non-linear with time.}
\label{fig:schema_evol_gamma} 
\end{figure}

The fact that PLT is observed to work very well up to close
to the time of NN domination, and the observation that 
Eq.(\ref{eq:omega1}) is valid when significant non-linear
correlation emerges, leads us to consider the approximation
of the early time evolution in which one abruptly matches
a PLT phase onto a NN dominated phase, i.e.,  

\begin{itemize}
\item {\bf Phase 1:} From $t=0$ up to a time $t_*$ particles in the
      system evolves according to PLT.

\item {\bf Phase 2:} For $t \geq t_*$ particles evolve only subject to
the gravitational attraction of their NN at the time $t_*$.

\end{itemize}

While the approximation used in the first phase is good for the whole
system, and in particular describes well the evolution of correlations 
at any distance, the second phase will only be valid approximately
in describing correlation at some sufficiently small scale, and 
for sufficiently short times. A schematic representation
of the evolution of correlations is given in 
Fig.~\ref{fig:schema_evol_gamma}.  
\begin{figure}
\includegraphics[width=0.5\textwidth]{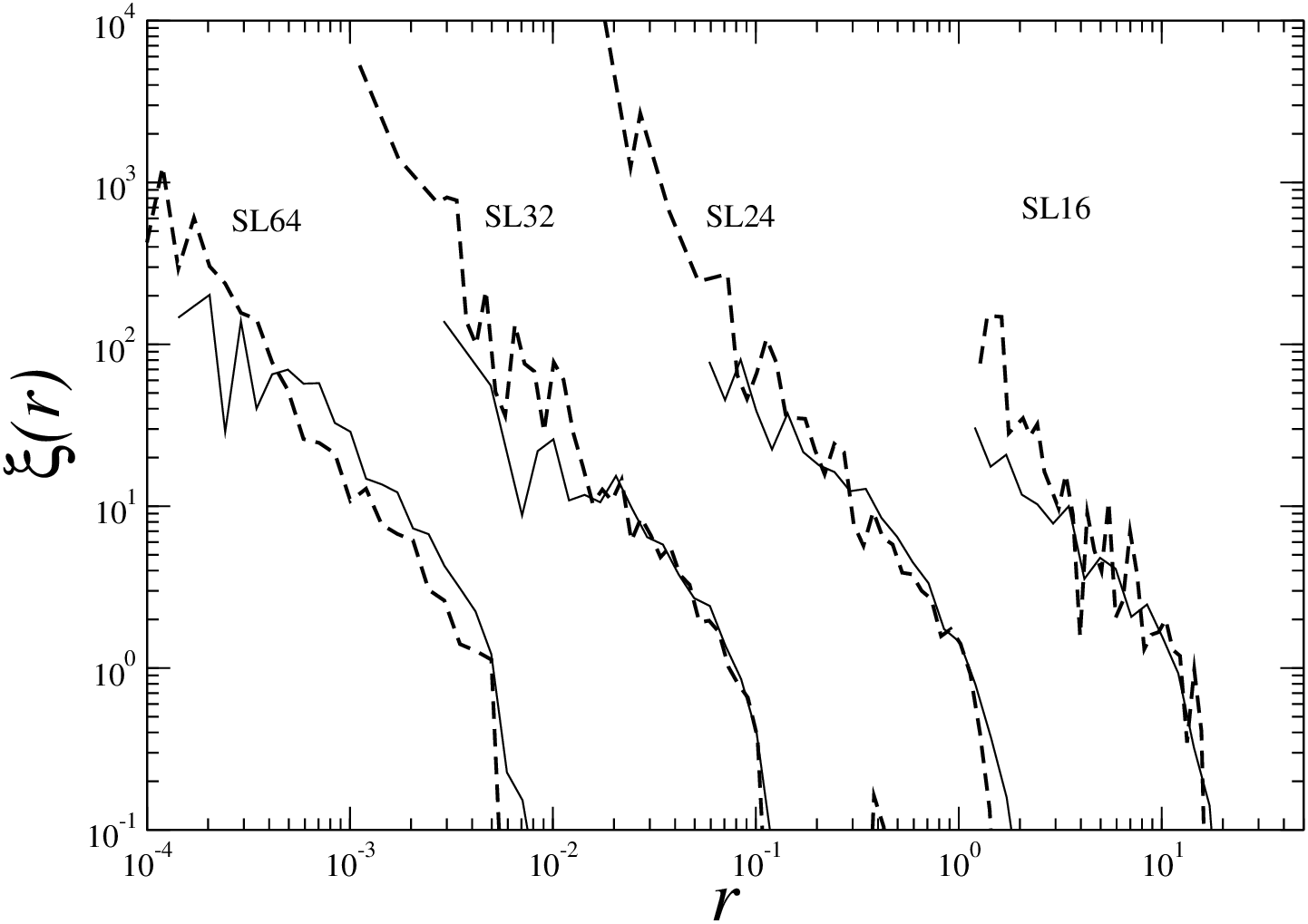}
\caption{
The two-point correlation function at the times
$t_\text{max}=1,2.5,3.5,4.5 \, \tdyn$ for the different initial 
conditions as indicated,  in both the full gravity simulations (thick lines) 
and the simulations of the two phase model described in the text
(thin lines). The corresponding transition times $t_*$ are the 
optimal ones given in Eq.~(\ref{eq:timeoftransition}). 
For clarity the $x$ axis has been rescaled for each initial
condition (as otherwise the curves are, to a very good
approximation, all superimposed, cf. \cite{sl1,sl2}). 
\label{fig:linear_NN}}
\end{figure}

We have implemented the above two phase evolution numerically on the
set of initial conditions given in Tab.~\ref{tab:sl_summary}. We have
taken the time at which we match the approximations, $t_*$, as a free
parameter and adjusted it to best fit the evolution of the
correlations in the full gravity simulations in the phase when 
strong non-linear correlation first emerge\footnote{In the 
second (NN) phase we use the same numerical value of 
$\varepsilon=0.00175$ as in the full
gravity simulations (and the same functional form of the
smoothing as in GADGET).}. We find that for each
initial condition there is indeed a choice of the time $t_*$ which 
gives such a fit, to a very good approximation over a range of amplitudes
from $\xi(r) \sim 10^2$ down to considerably less than unity. 
Results are shown in Fig.~\ref{fig:linear_NN}, for the optimal
times $t_*$: 
\begin{equation}
t_{*}\approx 0,\ 0.5,\ 1.5 \text{ and } 3.0 
\label{eq:timeoftransition}
\end{equation}
for SL64, SL32, SL24 and SL16 respectively (with time in units
of $\tdyn$). The results here are
given at the times $t_\text{max}$, which are the approximate 
times at which we observe the evolution under NN interactions
to lead to correlations beginning to deviate from (i.e. lag behind)
those in the full gravity simulations:
\begin{equation}
t_\text{max}=1, \ 2.5, \ 3.5  \text{ and } 4.5 \,.
\label{eq:maximaltime}
\end{equation} 
For times  
$t > t_\text{max}$ the modified simulations stop
evolving significantly (as one is then simply seeing the averaged
effect of the periodic motion of many NN pairs). In contrast
the full gravity simulations continue to evolve clustering 
from the collective motion of larger scales which has been
completely neglected in the second phase of the approximation.

\subsection{Transition to NN domination}

We now consider in more detail the essential time scale $t_*$, 
which we determined numerically above. Given our discussion of
and motivation for the two phase model, we might expect it to 
correspond to the time at which PLT breaks down and the forces 
on particles become typically dominated by that due to their NN. 
As we will now explain it corresponds in fact to a time somewhat 
shorter than this.

\begin{figure}
\includegraphics[width=0.5\textwidth]{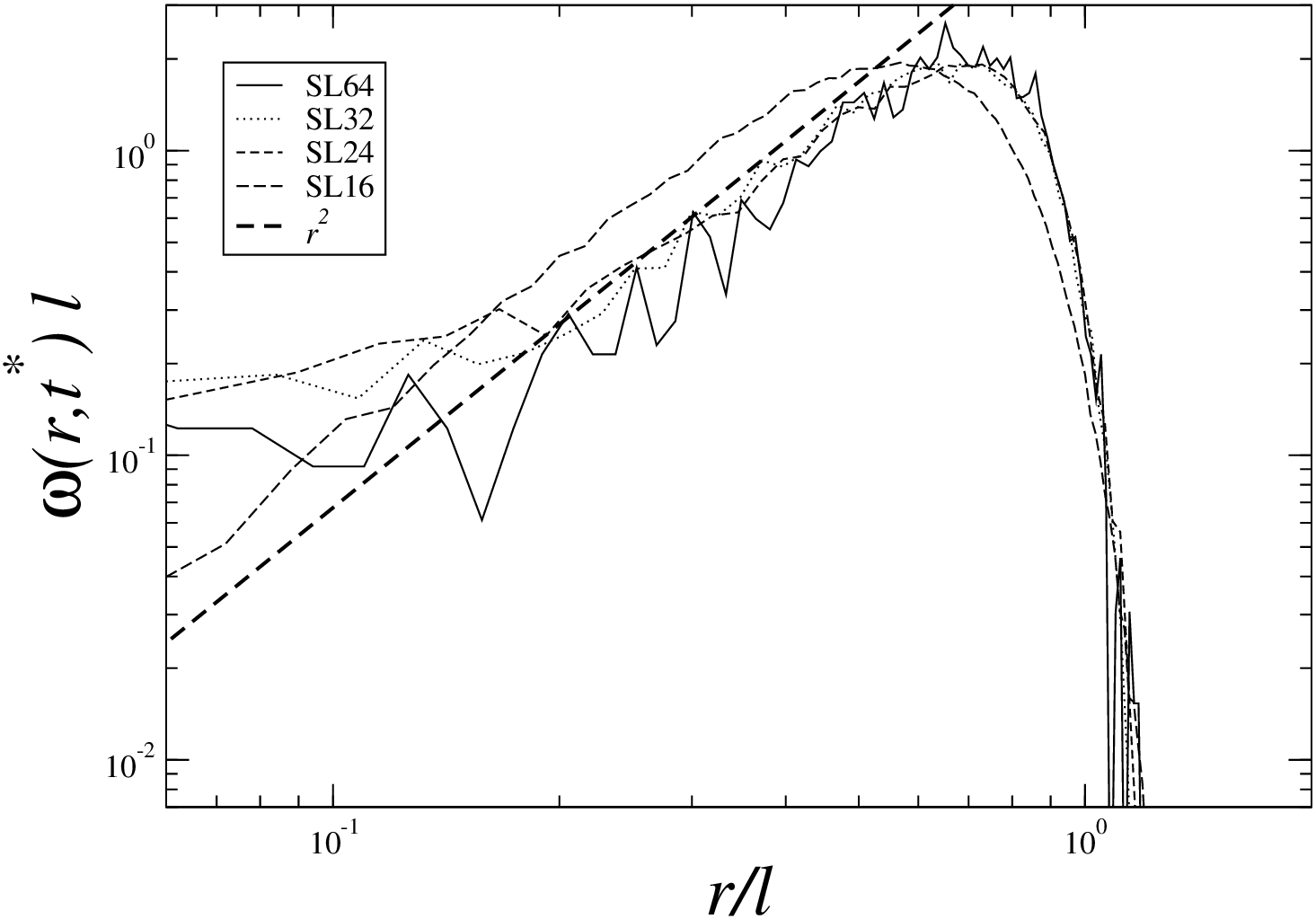}
\caption{The NN PDF in the distributions evolved to the times $t_{NN}$
given in Eq.~\ref{eq:tNN} for the different SL initial conditions
indicated.
Also shown is the result for a Poisson distribution 
[Eq.~(\ref{poisson-NNpdf})], and a line indicating
its small scale behavior $\omega_P(r) \propto r^2$.
Units on the $x$ ($y$) axis have been divided
(multiplied) by the numerical value of the lattice
spacing $\ell$ in each case.
\label{fig:nn_at_tNN}}
\end{figure}
\begin{figure}
\includegraphics[width=0.5\textwidth]{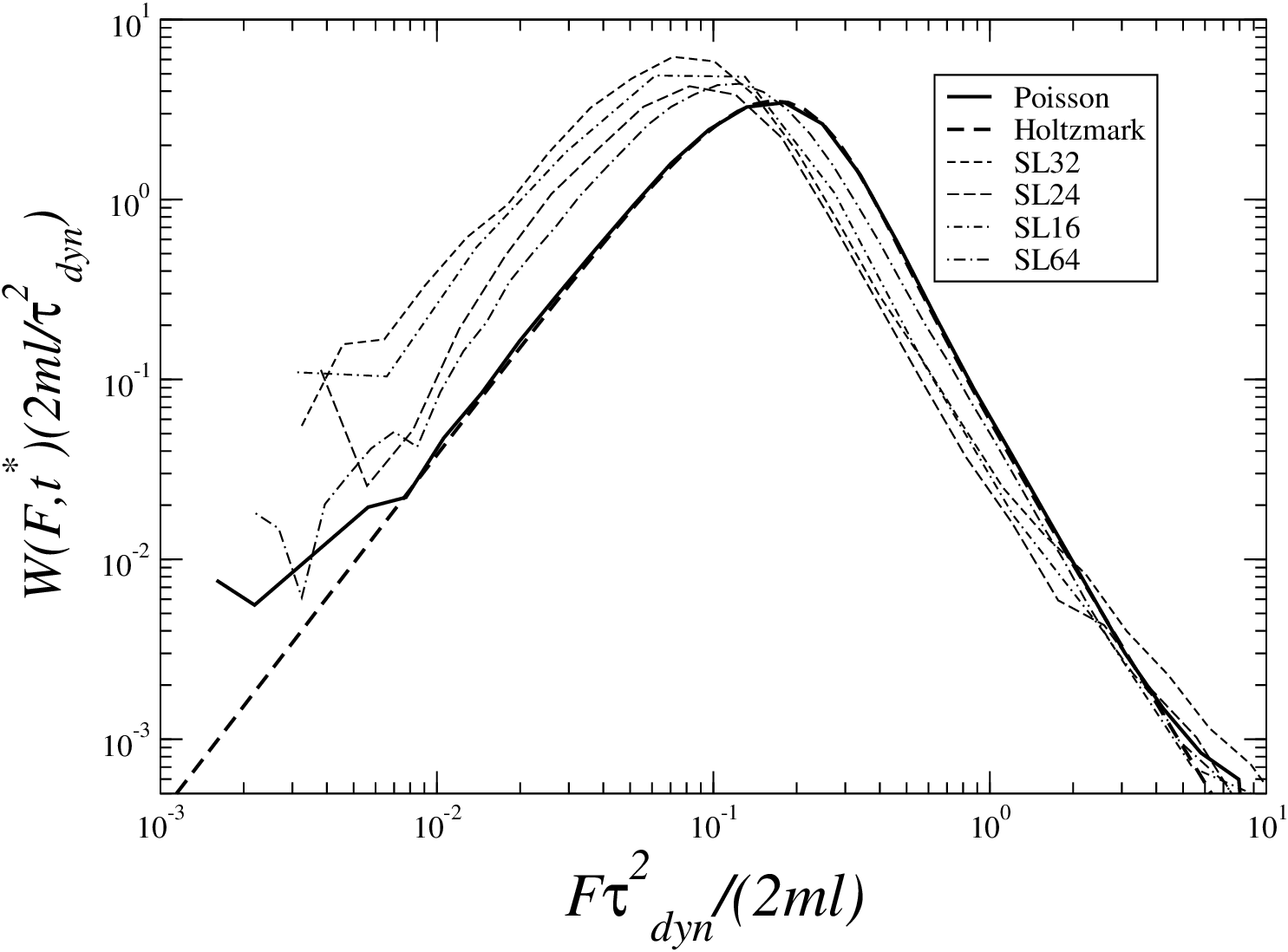}
\caption{PDF of the  force at the times 
$t_{NN}$ given in Eq.~\ref{eq:tNN} for the different SL initial 
conditions as indicated. The force has been rescaled as described 
in the text.  Also shown is the PDF for the 
P64 initial conditions, as well as the theoretical result for 
an infinite PDF (i.e. the  Holtzmark distribution).
\label{fig:wf_at_tNN}}
\end{figure}

An approximate characterization of the time $t_{NN}$ at which NN forces 
dominate can be given numerically by studying the relation between the
NN PDF $\omega(r)$ and the force PDF $W(F)$. As explained in Sec.~\ref{Force
distribution}, NN forces dominate when the relation 
Eq.~(\ref{NNdomination}) holds, for large values of the modulus
of the force $F$. While in the initial condition SL64 it already
holds to a very good approximation, in the distributions obtained by
evolution of the SL with smaller $\delta$ we find that it becomes
good at the times 
\begin{equation}
t_{NN} \approx 2.0,\ 3.0 \text{ and } 4.2 
\label{eq:tNN}
\end{equation}
for SL32, SL24 and SL16 respectively. Fig.~\ref{fig:nn_at_tNN} shows
the NN PDF in each case at these times, and Fig.~\ref{fig:wf_at_tNN}
the PDF of the force modulus. The quantities have been normalized by
the characteristic length/force scale in each case to give the PDF for
the corresponding dimensionless quantity: for the NN PDF we have
normalized the radial distance to the initial lattice spacing $\ell$
in each case, and for the force PDF we have normalized to the force
$(2 m \ell)/\tdyn^2 \propto Gm^2/\ell^2$ (i.e.  the force between two
particles at the characteristic distance $\ell$). We have plotted the
two quantities separately to show the fact that at the times $t_{NN}$
we find in all cases good agreement at small separations/strong forces
with the corresponding results for a Poisson distribution. This will
be important in our discussion below of the origin of the
``universality'' of the observed correlations.  As discussed in
Sec.~\ref{Force distribution} the relation Eq.~(\ref{NNdomination})
is very accurately followed for a Poisson distribution
(cf. Fig.~\ref{forceHNN}), and so it follows implicitly from these
figures that it is indeed a good approximation here in the regime of
strong forces. The very clear differences in the force PDF at smaller
values of the force are a reflection of the real difference in the
fluctuations at larger scales: the larger amplitude at smaller values
of the force, for a force PDF normalized as we have done, is a
reflection of the fact that the fluctuations become more suppressed at
large scales as $\delta$ decreases.  In the NN PDF, on the other hand,
such differences due to large scales are not present (because it
characterizes the small scale properties).

It is clear, from Eqs.~(\ref{eq:timeoftransition}) and (\ref{eq:tNN}),
that $t_*$ does not correspond, even approximately,
to $t_{NN}$, but rather is shorter by more than a dynamical time.
Likewise it is shorter than the estimated time of breakdown of
PLT. Indeed we find that at $t_{NN}$ we have in all cases that the
root square variance of the displacements of $NN$ particles, 
normalized in units of the appropriate $\ell$, is $\approx 0.5$
which corresponds to the criterion for breakdown of PLT found in
\cite{marcos_06}. At $t_*$, on the other hand, we measure 
the values
\begin{equation} 
\delta \approx 0.18, 0.12, 0.11 
\end{equation}
for SL32, SL24 and SL16 respectively\footnote{These values are, as
expected, also
in very good agreement with those predicted by PLT. A more appropriate 
characterisation of the breakdown of PLT is in fact given by considering
the variance of the {\it relative} displacements. The difference with
respect to the simple (one-point) variance is in fact negligible here 
for reasons that will be explained below.}.

This difference, and the approximate value of $t_*$, can in fact
be understood quite easily as follows. An approximation to the
evolution which is clearly better than our two phase approximation at 
all times is that in which we use the expansion of PLT to
approximate all the forces on a particle {\it except} that due to
the particle which becomes its NN in the second phase. Indeed 
we would expect such an approximation --- let is call it PLT+NN ---
to be very good for the whole regime we are treating. If we now
consider  our two phase treatment (PLT for $t<t_*$, NN for $t \ge t_*$),
as an approximation to PLT+NN, it is not difficult to understand
why the $t_*$ which makes the approximation optimal is of order 
a dynamical time smaller than $t_{NN}$. The reason is that 
the equation of motion for the displacements in PLT reduces,
for time scales up to of order a dynamical time, to a simple 
ballistic approximation\footnote{It is straightforward to show
that one obtains 
${\ve u} ({\ve R}, t)= {\ve u} ({\ve R}, 0) + {\ve v} ({\ve R}, 0)\,t$,
where ${\ve v} ({\ve R}, 0)$ is the velocity of the particle associated
with the lattice site $\ve R$ at the (arbitrary) initial time $t=0$, 
by expanding to linear order the full PLT expression for the evolution of
the displacements (see \cite{sl2}, or \cite{marcos_06}) in powers of 
$\epsilon_n (\ve k) t/\tdyn$ where $\epsilon_n (\ve k)$ are numbers of
order unity specifying the eigenvalues of the dynamical matrix for
gravity on the lattice \cite{marcos_06, sl2}. }. Thus when we turn on 
the NN interaction at time $t=t_*$ we follow, for of order a dynamical
time, the PLT+NN evolution. If we reduce $t_*$ 
further we will deviate from PLT+NN more because PLT is poorly
approximated; if we increase $t_*$ we will loose precision by
excluding the full NN contribution to the force.

Note that PLT ``comes for free'' in this way for a time after $t_*$
only because we match the velocities at $t_*$. A simple check on the
above explanation is provided by doing a simulation in which we reset
the velocities to zero at $t_*$. We have done this and find indeed
that the match to the full gravity evolution is considerably less
good, and that, particularly for the SL with smaller $\delta$ no
choice of $t_*$ can produce a good fit. We will return in the next
section to discuss further the role of the velocities.

\subsection{Origin of universality of non-linear correlations}

The simple two phase model thus describes quite well the emergence
of the non-linear correlations at early times for the
range of initial conditions studied. We now turn to the fact that
these correlations are approximately the same in all cases, i.e.,
the two point correlation functions, as a function of radial 
separation, agree quite well (and indeed agree well with these
quantities as found in \cite{Baertschiger:2004tx} and \cite{sl1,sl2}).

\subsubsection{Small scale correlation properties at $t_{NN}$}

The explanation for this ``universality'' of the clustering is clearly
suggested by the results shown above in Figs.~\ref{fig:nn_at_tNN} and
\ref{fig:wf_at_tNN}: in all cases the evolution gives at $t \approx
t_{NN}$ a point distribution with correlation properties very similar
to those of the Poisson distribution, at the scales relevant to the
development of clustering in the following phase.  Indeed the force
PDF, at stronger values of the force corresponding to the particles
which will cluster most rapidly in the subsequent time, follows very
closely that in the Poisson distribution. Thus the clustering then
develops as in the latter distribution giving the same correlation
properties. These are simply those which emerge, as described in
detail in \cite{Baertschiger:2004tx}, when pairs of particles with
initial separations given by the NN PDF $\omega_P(r)$ in the Poisson
distribution fall on one another.
   
Why are the correlation properties so similar at these scales
to those of a Poisson distribution at the time $t_{NN}$? We 
have emphasized in Sec.~\ref{sec2} that a Poisson distribution
is approximated to increasingly large scales as the amplitude
$\delta$ of the random {\it uncorrelated} shuffling increases.
It follows that if the evolution described by PLT is to a good 
approximation simply an amplification of the initial 
displacements, with only very weak correlation, the transformation
to a ``universal'' Poisson initial condition at the relevant scales
at $t \approx t_{NN}$ results. To quantify whether this is indeed
the case we can consider, e.g.,   
\begin{equation}
c_{NN} (t) \equiv \frac{\frac{1}{6} \sum_{{\ve R}_{NN}} 
\langle {\ve u} (0) \cdot {\ve u} ({\ve R}_{NN}) \rangle}
{\langle {\ve u}^2 \rangle}
\label{correlation}
\end{equation}
where ${\ve R}_{NN}$ are the lattice vectors of the six 
particles of the simple cubic lattice closest to the origin
(where there is a particle). The ensemble average is the 
average over realizations of the random initial conditions of
the SL\footnote{When this average is performed all six approximate
$NN$ are equivalent so that it is in fact sufficient to evaluate
$c_{NN}(t)$ for a single neighbor [and drop the sum
and factor of $1/6$ in Eq.~(\ref{correlation})].}.

\begin{figure}
\includegraphics[width=0.5\textwidth]{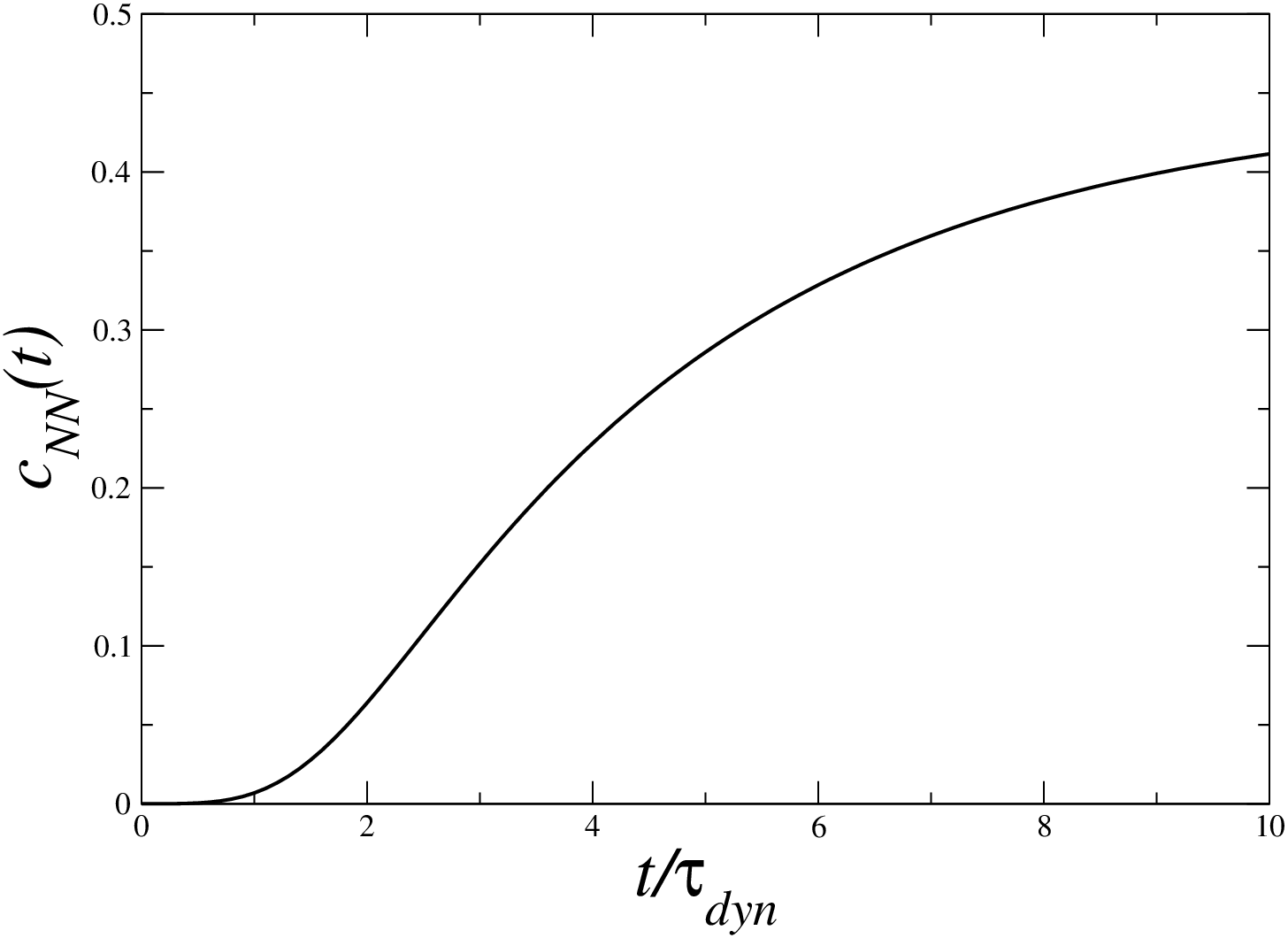}
\caption{Temporal evolution of the simple measure $c_{NN}(t)$ of the 
correlation in the displacements given by Eq.~(\ref{correlation}) in PLT.   
\label{fig:correlation}}
\end{figure}

In Fig.~\ref{fig:correlation} is shown $c_{NN}(t)$ as calculated
exactly in PLT. We see that, for the time scales over which we use
PLT (at most about four dynamical times), the correlation which 
develops between displacements at the relevant (small) scales is 
indeed weak ($c_{NN} < 20\%$). 

\subsubsection{The limit $\delta \rightarrow 0$}

Fig.~\ref{fig:correlation} shows that if, instead, our initial
conditions had $\delta$ sufficiently small so that $t_{NN}$ were  
greater than a few dynamical times, the approximation 
of weak correlation of the displacements at small scales would
become progressively worse as $\delta$ decreases.
As a result the basis for the approximate universality in 
the subsequent evolution would also be expected to become a
progressively poorer approximation. The evolution of the 
displacements in PLT
is simply a sum over the appropriately evolved eigenmodes of the
displacements fields in the corresponding linear approximation to
the inter-particle force. The behavior we observe here at long times
is a result, as discussed in detail in 
\cite{joyce_05, marcos_06, sl2}, of the fact that in this regime the
small spread in the eigenvalues of the modes of the displacement field
becomes important. The modes with slightly faster
growth become arbitrarily dominant, leading to the very specific
correlation of displacements described by these modes.
For arbitrarily long times, i.e., for arbitrarily small initial
$\delta$, one therefore obtains a distribution with correlation 
properties at all scales very different to that which could 
form from the Poisson distribution\footnote{Specifically, 
as $\delta \rightarrow 0$ the evolution
will always be dominated at the time when PLT breaks down 
by the most rapidly growing eigenmode. In an SL lattice 
(see \cite{joyce_05, marcos_06}) this eigenmode is one in 
which adjacent infinite parallel planes fall towards one 
another.}.
We thus conclude that the ``universality'' we observe in our 
numerical simulations is a good approximation in the range of
SL initial conditions with small, but not very 
small\footnote{Such small initial $\delta$ are very difficult to
simulate numerically because of the precision required.}, $\delta$.

\subsubsection{Role of velocities at $t_{NN}$}

\begin{figure}
\includegraphics[width=0.5\textwidth]{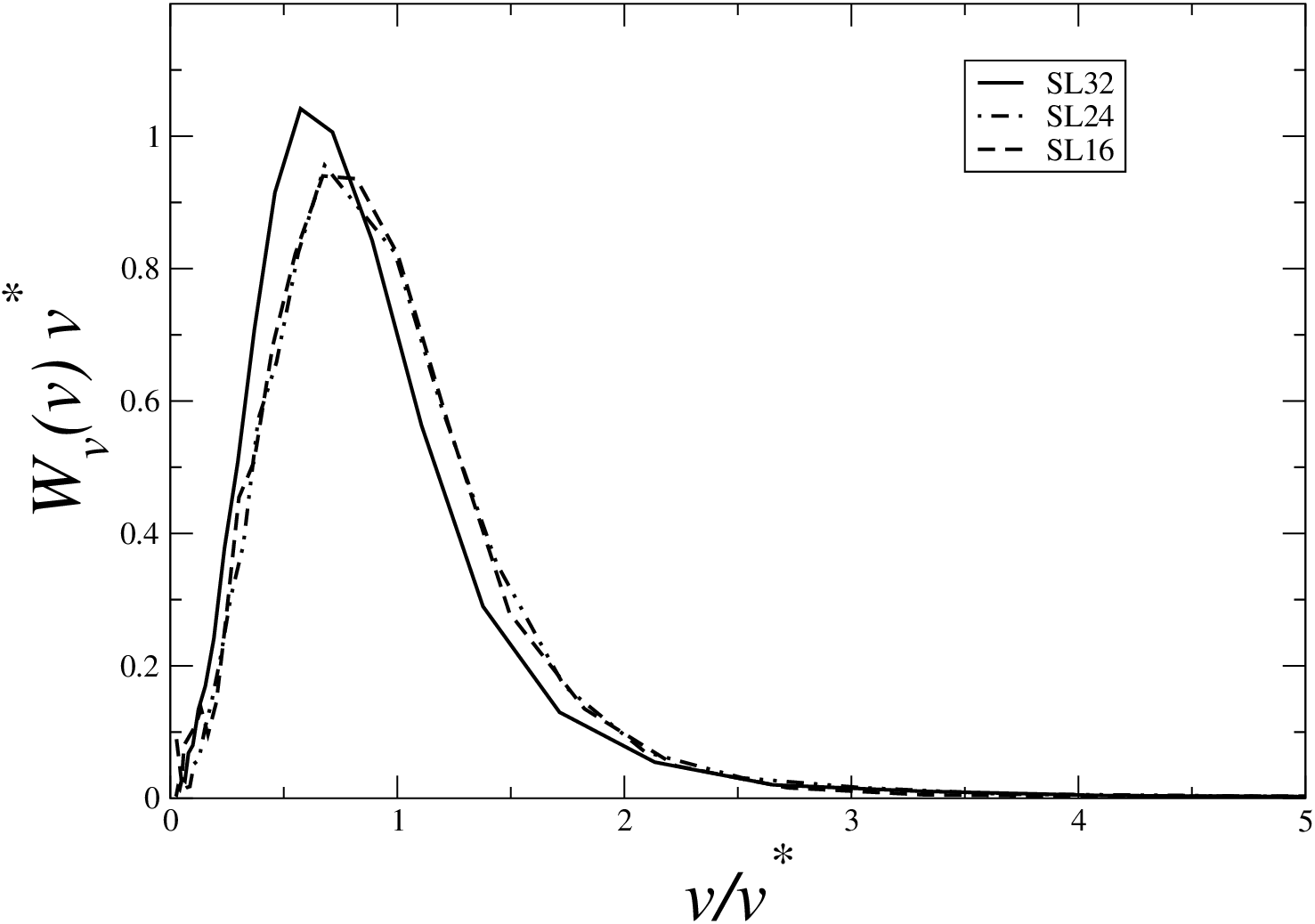}
\caption{PDF of the modulus of the velocities at time $t_{NN}$
in the full gravity simulations starting from the initial conditions
SL32, SL24 and SL16. The characteristic velocity $\overline{v}$ is
defined in Eq.~(\ref{vbar-defn}).}
\label{fig:vel_at_t*}
\end{figure}

In the above discussion we have neglected the role of the velocities:
in SL32, SL24 and SL16 non-zero velocities have developed at $t_{NN}$ 
which make the full initial conditions at this time different from
those in SL64 and P64 (with vanishing velocities at $t_{NN}=0$).
The PDF $W_v(v)$ of the modulus of these velocities as measured in the 
different simulations at this time are shown in Fig.~\ref{fig:vel_at_t*}.
We have normalized for convenience in units of 
\be
\overline{v} \equiv \sqrt{\frac{Gm}{\ell}} \;,
\label{vbar-defn}
\ee 
which
is the velocity gained by a particle initially at rest when it reduces
its separation by one half to a particle initially at distance $\ell$.

We observe that the different PDF of the velocities agree quite well,
which means that the full (space and velocity) initial conditions 
at $t_{NN}$ are indeed very similar in these 
simulations. Compared to 
SL64 and P64, however, the difference in velocities is a priori
significant: their magnitude is not small, but of order unity, in 
the units chosen, which are characteristic for the next stage of
free fall of NN particles which leads to the correlations. Given 
that at this time $t_{NN}$ we expect the velocity of particles to be,
on average, oriented towards their NN (since $t_{NN}$ is significantly
larger than $t_*$), and that in the approximated 
Poisson distribution the average NN distance is $0.55 \ell$, the 
distribution at $t_{NN}$ should be well approximated as one 
in which pairs of NN fall on one 
another, but starting at an earlier time.

\begin{figure}
\includegraphics[width=0.5\textwidth]{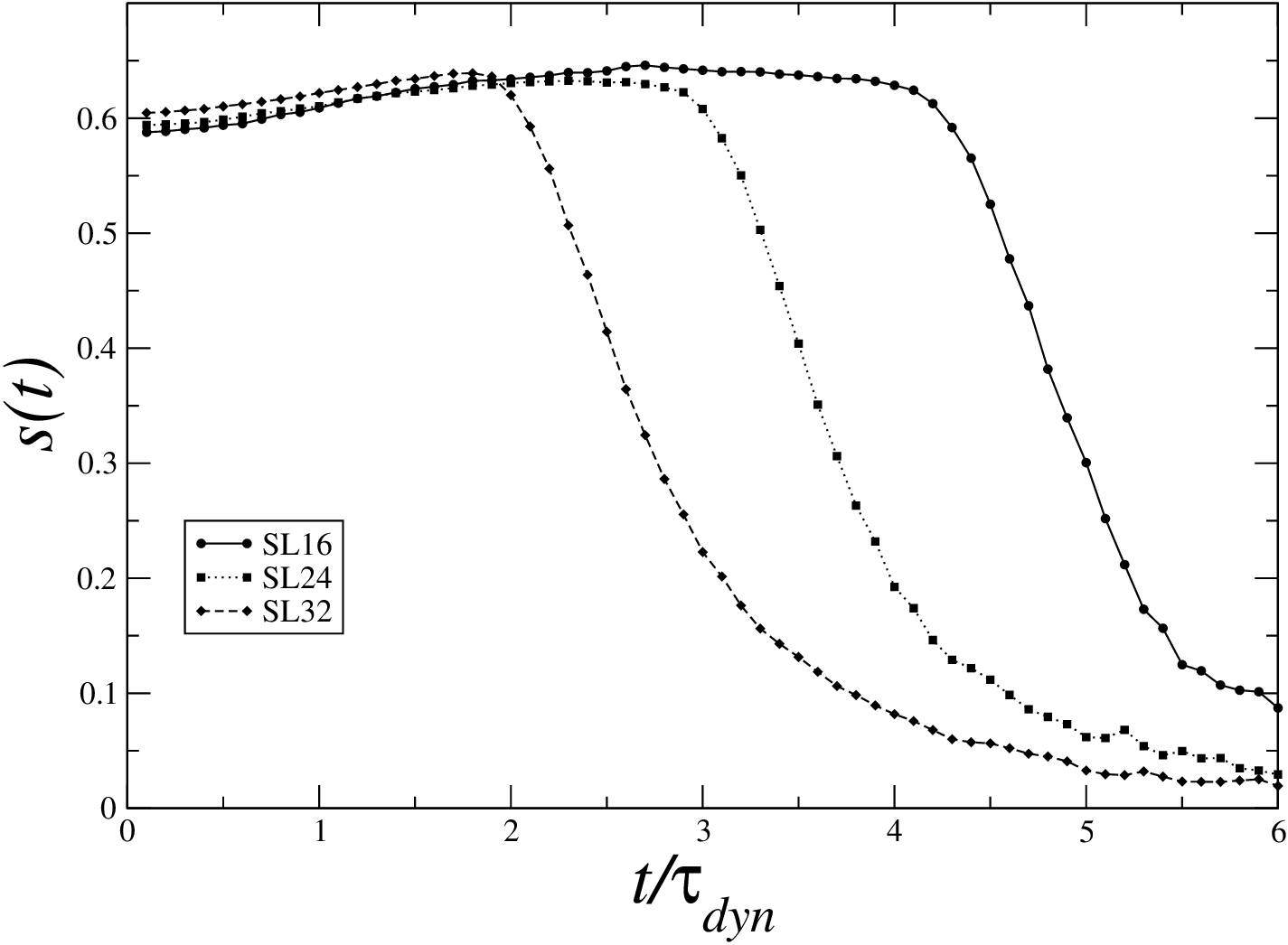}
\caption{Behavior of $s(t)$ [as defined in Eq.~(\ref{eqs})] in SL32, SL24 and SL16. }
\label{fig:s}
\end{figure}

This is further illustrated and quantified by Fig.~\ref{fig:s},
which shows the temporal evolution of the quantity
\be
\label{eqs} 
s (t) \equiv \frac{\langle {\ve v} \cdot {\ve r}_{NN} \rangle}{|{\ve v}| | {\ve r}_{NN}|}
\ee
for $t>0$, where $\ve v$ is the velocity of a given particle (at time $t>0$), 
and ${\ve r}_{NN}$ is the vector pointing towards its NN 
(at the same time $t$), and the average is over all the particles.
The evolution of this quantity is qualitatively 
very similar in all three simulations: after a 
slow rise from an initial non-zero value a rapid decrease
sets in at a time close to the estimated $t_{NN}$ in each 
case [cf. Eq.~(\ref{eq:tNN})]. The characteristic
time for this, roughly exponential, decay of the 
correlations corresponds well to 
$\Delta t=t_{\rm max}-t_{NN}$ [where $t_{\rm max}$ 
is as estimated in Eq.~(\ref{eq:maximaltime})].
We thus see, as anticipated, that there 
is significant correlation of the direction 
of velocity with the NN direction at time $t_{NN}$.
While $\Delta t \approx 1$ for SL64 and PL64, we have
$\Delta t \approx 0.5$ for the three simulations shown
here, with a very similar behaviour of the correlation
function $s(t)$ in this phase. Thus SL32, SL24 and SL16 
all lead to similar non-linear correlations as those 
in SL64 and P64, but in a shorter time due to this
correlation of velocities with the NN direction acquired
before $t_{NN}$.

The behavior of $s(t)$ observed here can be understood
in greater detail in the model we have described. The 
initial non-zero, approximately constant, value is a 
result of the fact that at sufficiently small times PLT can be
well approximated by its fluid limit. In this case
\cite{sl2, marcos_06} the displacement of each particle
off its lattice site is simply amplified in time,  
so that the function $s(t)$ is independent of time
and equal to the expression in Eq.~(\ref{eqs})
with ${\ve v}$ replaced by ${\ve u}$, the initial 
displacement of the particle from its lattice site 
(giving $s(0) \approx 0.6$).
The slow increase of correlation is due to the difference
between PLT and this fluid limit. The decrease from about
$t_{NN}$ signals that pairs of NN particles have now begun
to cross one another, giving a contribution with the opposite
sign to $s(t)$. At sufficiently long times a given 
particle's motion is finally no longer oriented with the
direction of its NN, as expected since the gravitational
field will become dominated by the collective effect
of many particles acting on any given particle.

\section{Discussion} 
\label{sec4} 

In this article we have studied in detail the early time
evolution of infinite self-gravitating shuffled lattices,
as well as the limiting case given by the Poisson distribution.
We have shown that a very good description of the evolution
of two point correlations in this phase is given by 
a simple approximation in which the force on particles
abruptly switches from that given by the PLT approximation
developed in \cite{joyce_05, marcos_06} to the force due
only to NN particles. Further in the first phase the system
evolves at small scales to always resemble closely the Poisson 
distribution, explaining the universality of the form of
the non-linear correlation function which emerges. We have noted,
however, that this universality will not extend to SL initial
conditions with arbitrarily small initial shuffling.
In this limit effects come into play in the very long first (PLT)
phase leading to a strongly correlated evolution at all scales,
which is different from (and unrelated to) that in the 
Poisson distribution.

We have thus given, for this specific class of initial conditions, an 
explanation of the non-linear correlations which emerge at early times.
As underlined in \cite{sl1,sl2} this non-linear correlation function
coincides with that which is observed at later times, when the system
manifests a simple spatio-temporal scaling (or ``self-similarity'').
Thus the model appears to explain this asymptotic form of the
non-linear correlations. As described in \cite{sl2} this can be
understood also in the following way: these non-linear correlations
in the system evolving at any later time can be well approximated   
by those in an evolved coarse-grained ``daughter'' distribution.
In the latter the system may be in the early time phase studied
here, while the original distribution is not. The non-linear
correlation functions of the two systems nevertheless coincide.
 
Our results are of relevance to simulations of structure formation in
the universe in cosmology. In this case the goal of numerical
simulation is to recover the non-linear correlations in the
Vlasov-Poisson (VP) limit of the evolution of a self-gravitating
$N$-body system. These initial conditions are different to those used
here --- with initially {\it correlated} displacements and an expanding
space --- but, as shown in \cite{sl1,sl2}, the evolution is
qualitatively similar to our simpler case. {\it If} the same kind of
model can be used to explain non-linear correlations in an early time
regime in this case then these correlations clearly are not described
by the VP limit: the forces due to NN particles are neglected in this
(mean field) limit, while in this model they are the dominant ones.
In this context this would mean that results in this regime would need
to be discarded as unphysical (i.e. not representing the required
physical limit, but just a numerical effect arising from the method of
discretization). Further, if the form of the asymptotic non-linear
correlation function is really determined by this early time
evolution, these effects of discreteness are then important at all
times and the system never represents the VP limit. This resemblance
of the early time and asymptotic correlation function may, however, be
a simple coincidence. The evolution at longer times may then indeed
be representative of the VP limit while discreteness effects are
important only at early times.

We will study these issues further in future work. Firstly the
application of our simple two phase description to cosmological
simulations should be investigated. We expect the expansion of
space to change nothing qualitatively in our model: the 
description of the PLT phase is qualitatively unchanged
\cite{marcos_06}, and NN domination during the formation
of the first non-linear structures in such simulations has 
been explicitly shown in \cite{Baertchiger:2002tk}.
The fact that 
the typical initial conditions of such
simulations have much more long wavelength power than those 
considered here may, however, be important [typically one has 
$P(k) \sim k^{m}$ with $m\leq -1$,
compared to $n=2$ (SL)and $n=0$ (Poisson) considered here]. Secondly,
as we have discussed at some length in the conclusions of \cite{sl2},
it would be instructive to study numerically the relation of the
asymptotic regime to the early time regime by completely modifying the
former using a large smoothing in the force, i.e., a smoothing scale
$\varepsilon \gg \ell$. In this case the dynamics we have described,
between $NN$ particles, will not occur. However at longer times the
system should still evolve the same correlations if the VP limit of
the system is indeed that represented by the simulations with
$\varepsilon \ll \ell$.

\acknowledgments{
We thank the ``Centro Ricerche e Studi E. Fermi'' (Rome) for the use
of a super-computer for numerical calculations and the MIUR-PRIN05
project on ``Dynamics and thermodynamics of systems with long range
interactions" for financial support.  We are grateful to Andrea Gabrielli
for many useful discussions.}


\end{document}